\RequirePackage[colorlinks,allcolors=blue]{hyperref}
\documentclass{agujournal2019}
\usepackage{url}
\usepackage{soul}
\usepackage{amsmath}
\usepackage{amssymb}
\usepackage{tablefootnote}
\journalname{arXiv}

\newcommand{\vu}{\mathbf u}
\newcommand{\vx}{\mathbf x}

\begin{document}

\title{Tidal fluctuations and spatial heterogeneity lead to trapping and chaotic mixing in coastal aquifers}
\authors{Satoshi Tajima\affil{1}, Marco Dentz\affil{2}}
\affiliation{1}{Graduate School of Frontier Sciences, The University of Tokyo, Tokyo, Japan}
\affiliation{2}{Spanish National Research Council (IDAEA-CSIC), Barcelona, Spain}
\correspondingauthor{Satoshi Tajima}{m@st-gw.net}

\begin{abstract}
The combined effect of tidal forcing and aquifer heterogeneity leads
to intricate transport patterns in coastal aquifers that impact both on
solute residence times and mixing dynamics. We study these
patterns through detailed numerical simulations of density-dependent
flow and transport in a three-dimensional heterogeneous coastal
aquifer under tidal forcing. Advective particle
tracking from both the freshwater and seawater domains reveals the
formation of chaotic and periodic orbits in the freshwater-saltwater
transition zone that may persistently trap contaminants.
We find that increasing heterogeneity results in increased trapping,
but also increased mixing entropy, which suggests that the chaotic
orbits enhance mixing between contaminants
from the freshwater and seawater domains. These findings highlight on
the one hand, the long-term contamination risks of coastal aquifers
through trapping, and on the other hand, the creation of hotspots for
chemical and biological reactions through chaotic mixing in the
transition zone.
\end{abstract}

\section{Introduction}
Coastal aquifers are characterised by dynamic interactions between seawater and
freshwater, where the interplay between tidal fluctuations, spatial
heterogeneity and aquifer compressibility play a significant role
\cite{werner2013seawater}. These processes
collectively govern flow dynamics, salinity distribution, and contaminant
transport within coastal aquifers. Density effects between seawater and
freshwater lead to the formation of a saltwater wedge that intrudes into the
coastal aquifer. Within the wedge, seawater recirculates from the seaward
boundary, while freshwater flows over the seawater body from the land
\cite{werner2013seawater}. At the interface between saltwater and freshwater, a
mixing or transition zone is created that represents a hotspot for chemical and
biological activity \cite{Sanford1989, Moore1999, Rezaei2005,
  Spiteri2008, de2020heterogeneity}.
These processes are subject to spatial heterogeneity in the hydraulic aquifer
properties and tidal fluctuations, which impact the
groundwater flow patterns, the dispersion of the transition zone, the mixing between salt and freshwater,
and the mixing and dispersion of contaminants in the aquifer. 

Studies in homogeneous and layered coastal aquifers have shown that tidal
fluctuations expand the transition zone between salt and freshwater
\cite{Oberdorfer1990, Inouchi1990, Pool2014}. Furthermore, \citeA{pool2015effects} showed
that spatial aquifer heterogeneity mitigates the impact of tidal fluctuations on the
width of the mixing zone between salt and freshwater, whereas increased aquifer
compressibility leads to an amplification. Using lithology of coastal
aquifers, \citeA{michael2016geologic} found that heterogeneity creates
complex spatial salinity distributions that lead to groundwater
circulation rates that cannot be captured by models based on
equivalent homogeneous media. For intertidal aquifers,
\citeA{geng2020heterogeneity} found that spatial heterogeneity creates
topological flow characteristics that encompass spatiotemporal
patterns and regions of intense and low mixing. Regions of strong mixing are amplified by spatial heterogeneity, which leads to the emergence of reaction hotspots along the saltwater-freshwater interface \cite{pool2018effects}, and facilitates the propagation of karst in coastal aquifers  \cite{de2020heterogeneity}.

The flow patterns in coastal aquifers induced by tidal fluctuations,
buoyancy and spatial heterogeneity also determine the transport, mixing and reaction behaviour of sea and land borne contaminants
\cite{robinson2018groundwater,santos2021submarine}. \citeA{geng2020numerical}
analysed solute transport in heterogeneous beach aquifers subject to
tidal sea level fluctuations. They found that spatial heterogeneity
enhances the spreading of a contaminant plume and the generation of
transient preferential flow paths and highly variable solute transit
times, and related these behaviours to the flow topology.
For constant density flows, the interplay
between temporal flow fluctuations, spatial heterogeneity and aquifer
compressibility can generate Lagrangian coherent structures and induce
chaotic advection and mixing \cite{trefry2019temporal, trefry2020lagrangian, wu2024lagrangian,
  tajima2024tide}. \citeA{trefry2019temporal} investigated advective transport in
horizontal two-dimensional heterogeneous confined aquifers under
temporal forcing superposed to a regional flow gradient. They find that heterogeneity and periodic temporal
forcing generates chaotic advection and Lagrangian coherent structures
near the forced boundary, which enhances fluid mixing and transport and
leads to anomalous residence time distributions, see also
\citeA{wu2020complex}. \citeA{tajima2024tide} analysed the spreading of
displacement fronts in two-dimensional heterogeneous aquifers driven
by transient forcing. They observe the creation of stable and chaotic
regions, which leads to containment and at the same time may promote
mixing.

In this paper, we investigate the
combined effect of buoyancy, tidal forcing and aquifer heterogeneity on flow and
transport patterns in three-dimensional, heterogeneous coastal aquifers. To this end, we analyze particle paths originating from the sea and land boundaries to
analyze the chaotic flow behaviour in terms of Poincar\'e
sections. The analysis of residence time distributions in the aquifer
elucidates further the creation of periodic and chaotic orbits,
quantifing how contaminants can be trapped in the transition zone
between seawater and freshwater. The evolution of the mixing entropy and segregation
intensity demonstrates how spatial heterogeneity leads to the mixing of sea and land borne solutes. 

\section{Methods}
\begin{figure}[htbp]
    \includegraphics[width=\linewidth, pagebox=artbox]{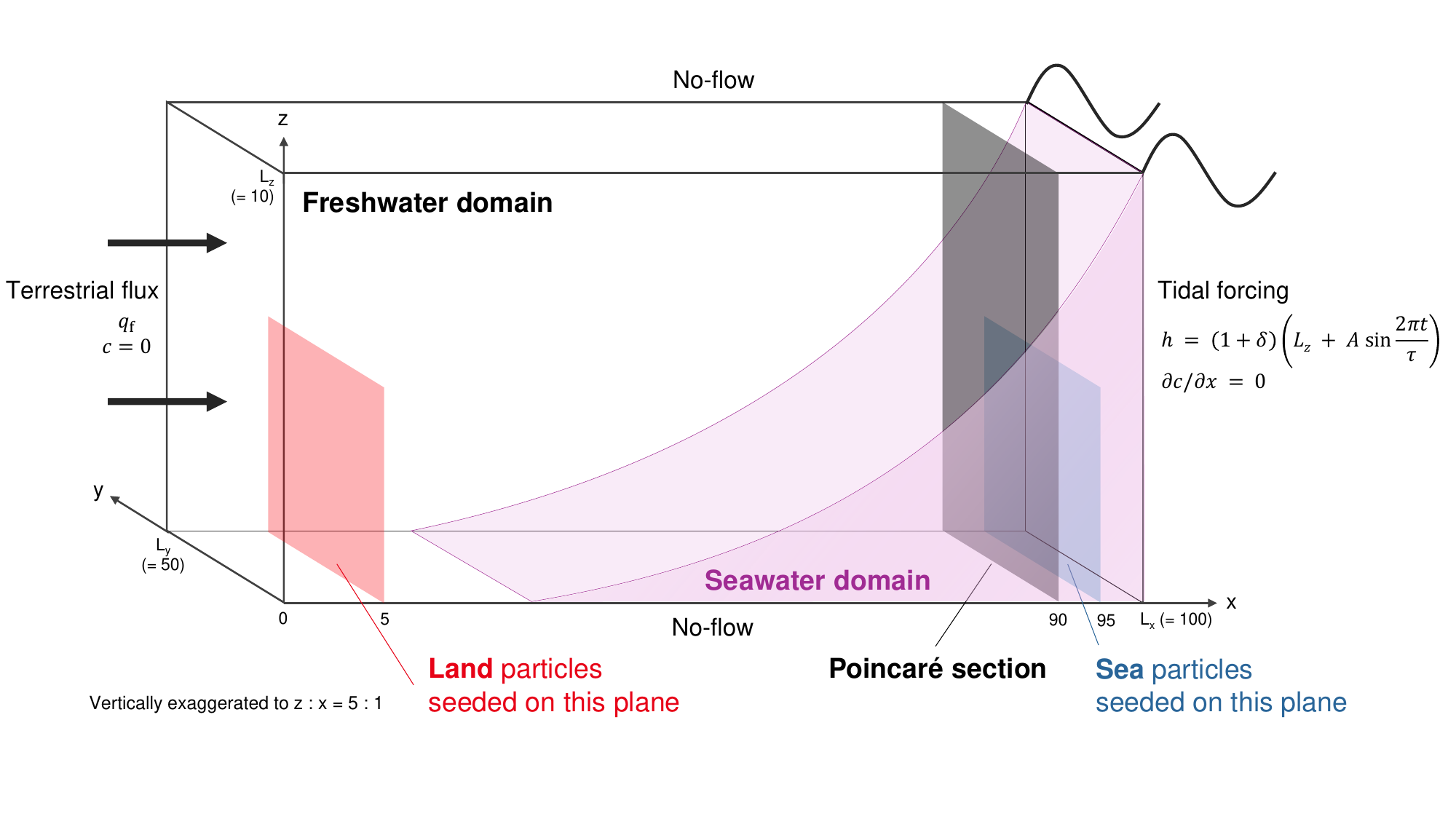}
    \caption{Schematic of model domain. Salinity distribution shown here is conceptual and not simulated result.}
    \label{fig:domain}
\end{figure}

We consider density-dependent flow in a fully saturated, three-dimensional, heterogeneous coastal aquifer
under tidal forcing. The setup and dimensions of the model domain are illustrated
in Figure \ref{fig:domain}. In the following, we outline the generation of the
spatially variable hydraulic conductivity field, the governing equations for
variable-density flow and particle tracking and their numerical solution, as
well as the observables that are used to elucidate the Lagrangian flow
kinematics, and the mixing potential.

\subsection{Hydraulic conductivity}
To systematically represent spatial aquifer heterogeneity, the hydraulic
conductivity, which is assumed scalar for simplicity, is
modelled as a multi-Gaussian random space function
\cite{Rubin-2003-Applied}. We consider the hydraulic conductivity $K_0(\vx)$,
which is referred to the freshwater density, that is, $K_0(\vx) =
k_0(\vx)/\rho_0 g$, where $\rho_0$ is the freshwater density, $g$ is
gravitational acceleration and $k(\vx)$ is permeability. Here, the
log-hydraulic conductivity $f(\vx) = \ln K_0(\vx)$ is modelled as a
correlated Gaussian random space function.  It has a constant
mean $\overline{f(\vx)} = \ln K_G$, where $K_G$ is the geometric mean
conductivity which is at the same time the effective conductivity
\cite{renard1997calculating}. Furthermore, we employ the exponential covariance
function 
\begin{equation}
\mathcal C_f(\vx) = \sigma_{f}^2 \exp\left[-\sqrt{(x^2 + y^2)/\lambda_\parallel^2 + z^2/\lambda_\perp^2} \right]
  \label{eq:Kcov}
\end{equation}
where $\sigma_{f}^2$ is the variance of $f(\vx)$, $\lambda_\parallel = 5$ m is the
correlation length in the $x-y$ plane, and $\lambda_\perp =
2$ m is the correlation length in the transverse direction. To quantify the effect of heterogeneity on the transport and
mixing behaviour, we consider $\sigma^2_f = 0.5, 1, 2, 4$. For each case, 20
realizations of the hydraulic conductivity are generated. A complete list of the modelling parameters is given in Appendix. 

\subsection{Variable-density groundwater flow}
The variable density groundwater flow model follows the formulation of
\citeA{Langevin2020}. Specifically, we use the Oberbeck-Boussinesq approximation,
which disregards density variations in the continuity equation. With this approximation, the
continuity equation is given by
\begin{align}
\label{eq:continuity}
S_s \frac{\partial h(\vx,t)}{\partial t} + \nabla \cdot \mathbf q(\vx,t) = 0,
\end{align}
where $S_s$ is the specific storativity, $h(\vx,t)$ is the hydraulic head, and $\mathbf
q(\vx,t)$ is the specific discharge. Momentum conservation is described by the Darcy
equation in hydraulic head form \cite{Langevin2020}
\begin{align}
\label{eq:Darcy}
\mathbf q(\vx,t) = - K_0(\vx) \left\{\frac{\rho(c)}{\rho_0} \nabla h(\vx,t) +
[h(\vx,t) - z] \frac{\nabla \rho(c)}{\rho_0}  \right\}. 
\end{align}
The fluid density $\rho(c)$ is assumed to depend linearly on
the salt concentration $c(\vx,t)$, that is, $\rho(c) = \rho_0 [1 + \epsilon
c(\vx,t)/c_s]$, where $c_s$ is the constant salt concentration in seawater, and
$\rho_0$ is the reference density of freshwater. The evolution of the salt
concentration $c(\vx,t)$ is described by the advection-dispersion equation
\begin{align}
\label{eq:c}
\frac{\partial c(\vx,t)}{\partial t} + \nabla \cdot \left[\mathbf u(\vx,t) c(\vx,t) \right] -
\nabla \cdot \left[ \mathbf D(\vx,t) \nabla c(\vx,t) \right] = 0,
\end{align}
where $\mathbf u(\vx,t) = \mathbf q(\vx,t)/\phi$ is the porewater velocity,
and $\phi$ is the porosity. The local hydrodynamic dispersion tensor $\mathbf{D}(\vx,t)$
is given by~\cite{Bear1972}
\begin{align}
    \label{eq:D_def}
    D_{ij}(\vx,t) = D_m \delta_{ij} + \delta_{ij} \alpha_I |\vu(\vx,t)| +
    (\alpha_I - \alpha_{II}) \frac{u_i(\vx,t)u_j(\vx,t)}{|\vu(\vx,t)|},
\end{align}
where $D_m$ is the molecular diffusion, $\alpha_I$ and $\alpha_{II}$ are the longitudinal and transverse dispersivity, respectively.
Terrestrial freshwater flux $q_\mathrm{f}$ with zero salinity
($c=0$) is applied to the inland lateral boundary ($x=0$). The sea boundary
($x=L_x$) is forced with tides, expressed by sinusoidal fluctuations in hydraulic head as
\begin{align}
    \left.h(\vx,t)\right|_{x = L_x} = L_z + A \sin\frac{2\pi t}{\tau} \label{eq:head_BC}, 
\end{align}
where $t$ is the time, $A$ is the amplitude, and $\tau$ is
the period. Here, we assume a semidiurnal ($\tau = 0.5$ d) tide with an
amplitude of $A = 0.5$ m \cite{schwiderski1980charting, Pool2014,
  pool2015effects, schrama1994preliminary}. Using these values for tidal
properties and a specific discharge of $S_s=1\times10^{-2}$ m $^{-1}$, the tidal
mixing number \cite{Pool2014} is $n_{\mathrm{tm}}\equiv\sqrt{\tau K_G/S_s}/A
\approx 26.5 \leq 600$, indicating the potential for significant tidal effects
on mixing and spreading \cite{pool2015effects}. These boundary conditions create
mean flow from the inland boundary at $x=0$ towards the coastal boundary at
$x=L_x$, aligned with the $x-$ and $z-$ axes and perpendicular to the
$y$-axis. The coastal boundary is assumed to be a no-dispersion boundary
($\partial c / \partial x=0$). This means that, if inflow from the sea to the
model domain occurs, the concentration at the boundary is identical to the
normalised salinity of seawater ($c=1$) \cite<e.g.>{Pool2014,
  tajima2022estimating, tajima2024mechanisms}. In contrast, if outflow from the
domain occurs, the boundary concentration becomes identical to that of the
discharging groundwater. The other boundaries are assumed to be no-flow. The
boundary conditions are summarised in Figure \ref{fig:domain}, and the modelling
parameters are summarized in Appendix. 

The numerical simulations based on the density-dependent flow and transport system
given by equations~\eqref{eq:continuity}--\eqref{eq:c} are performed with
MODFLOW 6~\cite{Langevin2017}. The model domain is spatially discretised so
that the mesh P\'eclet number is below 2 throughout the domain to ensure
convergence. We first perform simulations without tidal fluctuations until a
quasi-steady state. Tidal fluctuations are then included, and the simulations
are run until a dynamic steady state is reached, where the temporal
fluctuations in the head and salinity distributions become stable.

\subsection{Particle tracking}
To explore the impact of tidal fluctuations,
heterogeneity and variable density on solute transport, we analyze advective
particle trajectories originating from the sea and land boundaries. We focus on the
distribution of residence times in the system and the advective interpenetration of
particle trajectories in the seawater- and freshwater-dominated aquifer
regions. Advective particle tracking elucidates the Lagrangian kinematics and
transport structure of the flow \cite{wu2020complex}. The trajectory of a particle is
given by the kinematic equation~\cite{kubo2012statistical},
\begin{equation}
    \frac{d \vx(t,\mathbf{a})}{dt} = \mathbf{u}[\vx(t,\mathbf{a}),t],
    \label{eq:Langevin}
\end{equation}
where $\vx(t,\mathbf{a})$ is the particle position at time $t$ with origin at
$\vx(t=0,\mathbf{a}) = \mathbf{a}$. We consider two families of particles with different initial positions at
$t=0$. Each family is vertically aligned on a single $y-z$ cross section within
$2.5 \leq y \leq 47.5$ m and $1 \leq z \leq 5$ m. The ''sea'' family (coloured in red) is located at a plane at $x=5$ m, whereas the ''land'' family (coloured in
blue) at $x=95$ m as illustrated in Figure \ref{fig:domain}. 
Equation \ref{eq:Langevin} is numerically solved using a
third-order Runge-Kutta method \cite{tajima2024ted, tajima2024tide}.
Further details on the setup of the numerical particle tracking simulations are
given in Appendix.

\subsection{Observables}
From the simulated particle trajectories, we investigate the Lagrangian
kinematics and residence time distributions of the flows. The evaluation of
residence times gives insight in whether there are closed orbits in which
particles can get trapped. To analyze the types of flows that may
occur, we also determine Poincar\'e sections. To assess the
mixing of particles originating from the sea and land boundaries, we
consider the mixing entropy and segregation index as outlined below.  

\subsubsection{Residence time distributions}
The residence time distributions from the land and sea boundaries are denoted by
$R_L(t)$ and $R_S(t)$, respectively. We consider the cumulative distributions
$R_i(t) \equiv n_i(t)/n_{0,i}$ ($i = L, S$), where $n_i(t)$ is
the number of particles from the land ($i = L$) and sea ($i = S$) boundaries
that have discharged over the sea boundary before time $t$, and $n_{0,i}$ the
respective total number of particles. Thus, $R_i(t)$ is equal to the fraction of
particles that have discharged from the domain by time $t$ relative to the
initial number of particles, or, equivalently, the probability that the
residence time is smaller than $t$. These observables are determined from the
respective advective travel times to the sea boundary. That is, for $R_L(t)$, the
travel time from the land to the sea boundary, and for $R_S(t)$, the
travel time from the sea to the sea boundary. The trapping of particles along
periodic or aperiodic orbits leads to infinite residence times, which manifests
in asymptotic values of $R_i(t)$ smaller than $1$.  

\subsubsection{Poincar\'e sections}
To analyze the Lagrangian kinematics and transport structure of the flow driven
by the interplay of tidal forcing and aquifer heterogeneity, we investigate 
Poincar\'e sections. While three-dimensional particle
trajectories may be complex and are in general more diffusive to visualize,
Poincar\'e sections represent the distribution of intersection
points of advective particle trajectories on a two-dimensional plane, oriented
perpendicular to the mean flow and aligned with the $y-$ and $z-$ axes as
illustrated in Figure \ref{fig:domain}. The Poincar\'e section represents a
mapping of the plane onto itself. The times between intersections are generally
not constant \cite{strogatz2001nonlinear}. Periodic orbits intersect the plane always at
the same point, quasi-periodic orbits at a finite set of points or along a closed
curve in the plane. Non-periodic orbits intersect the plane at each iteration at a
different point, which indicates chaotic mixing. Here we consider two families
of particles according to their origin at the sea and land boundaries
marked by blue and red, respectively. For tidally-forced variable density flow
in a homogeneous aquifer, there are no recurring trajectories, that is, each
trajectory intersects the target plane only once. Therefore, the number of
intersections remains constant in time. As we will see in the following, this is different for heterogeneous aquifers; we will observe both periodic and
aperiodic orbits, which are commensurate with the trapping of particles.
Non-periodic orbits lead to an increase of the number of intersection points in the
plane with the number of iterations. Note that the Poincar\'e sections considered
here are different from the ones considered in \citeA{wu2024lagrangian} and
\citeA{tajima2024tide} for two-dimensional constant density flow in periodically
forced aquifers. These authors define Poincar\'e section as a stroboscopic map capturing the locations of advected particles at each flow period. In other
words, the Poincaré sections in these works generate a lower-dimensional
subspace in a temporal domain by sampling particle positions at discrete
time intervals corresponding to tidal periods. In contrast, the Poincar\'e sections
considered in this work record intersection points on a two-dimensional plane
from three-dimensional particle trajectories.

\subsubsection{Mixing entropy}
In order to assess the potential of the two families of particles to mix, we
use the concept of mixing entropy. Specifically, to determine the degree of mixing between
the two families and among each family separately, we apply the Shannon entropy
measure proposed by \citeA{camesasca2006quantifying} to the set of points or
dots of different colour on the Poincar\'e sections at $x=90$ m. To do so, we
discretize the plane into a square grid of grid length $\Delta x$. Then, we define the joint probability $p_{j,k}$ that a point
on the Poincar\'e section belongs to family $j$ ($j = S$ denotes the blue
particles from the sea boundary, and $j = L$ the red particles from the land) and
resides in bin $k$,
\begin{align}
    p_{j,k} = \frac{n_{j,k}}{\sum_{j,k} n_{j,k}},
\end{align}
where $n_{j,k}$ is the cumulative number of intersection points of the family
$j$ in bin $k$. The joint distribution $p_{j,k}$ can be expanded using the Bayes
formula as $p_{j,k} = p_{j|k} p_k$, where
\begin{align}
    p_{j|k} = \frac{n_{j,k}}{\sum_j n_{j,k}}, && p_k = \frac{n_k}{\sum_k n_k}, && n_k = \sum_j n_{j,k}. 
\end{align}
That is, $p_{j|k}$ is the probability that an intersection point that is in bin $k$ belongs to family $j$, and $p_k$ is the probability of finding an intersection point in bin $k$. The
entropy $H$ of the distribution of intersection points is then
defined by
\begin{align}
\label{eq:H}
    H = - \sum\limits_{j,k} p_{j,k} \ln(p_{j,k}). 
\end{align}
Using the decomposition $p_{j,k} = p_{j|k} p_k$, it can be written as $H = H_c +
H_s$, where $H_c$ denotes the mixing entropy between the two families, or
colours, and $H_s$ of the distribution of intersection points irrespective of the
colour,
\begin{align}
    H_c = - \sum_k \left[p_k \sum_j p_{j|k} \ln(p_{j|k})\right], && H_s = - \sum_k p_k \ln(p_k). 
\end{align}

\subsubsection{Segregation intensity}
The segregation intensity of \citeA{danckwerts1952definition} measures, as the
name implies, the degree of segregation of the two colours (families). It is complementary
to the mixing entropy defined in the previous section. We consider the same
setup as in the previous section and want to determine the degree of segregation
of the intersection points of different colours in the Poincar\'e section. Thus,
the segregation intensity $I$ is defined as
\begin{align}
    I = \frac{\sigma_r \sigma_b}{\overline p_r \overline p_b}. 
\end{align}
The average probability $\overline p_j$  of group $j$ in the Poincar\'e section is
\begin{align}
    \overline p_j = \frac{1}{N} \sum_{k \in E} p_{j|k},
    \label{eq:conc_ave}
\end{align}
with $N$ is the total number of bins and $E = \{k|p_k \neq 0\}$ . The standard deviation $\sigma_j$ of the
distribution of $j$ is defined by
\begin{align}
    \sigma_j^2 = \frac{1}{N} \sum_{k \in E} \left(p_{j|k} - \overline p_j\right)^2.
    \label{eq:conc_sd}
\end{align}
For illustration, let us consider the situation of full segregation, as is the
case for a homogeneous aquifer. The red dots of the land family are confined
to a fraction $\rho_r$ of the Poincar\'e section with constant density $a_r$
and the blue dots of the sea family to the complement with fraction $\rho_b
= 1-\rho_r$ and constant density $a_b$. Then, the mean probability and variance are 
$\overline p_j = \rho_j a_j$ and  $\sigma_j^2 = a_j^2 \rho_j (1 - \rho_j)$.
In this case of full segregation, the segregation intensity is $I = 1$. In the
case of full homogeneity, the distributions of red and blue dots are $p_j =
\rho_j a_j$ with $j = r,b$. In this case, $\overline p_j = p_j$, $\sigma_j^2 =
0$, and the segregation intensity is $I = 0$.

\section{Results and discussion}
\subsection{Emergence of confined pathlines and trapping}
\begin{figure}[htbp]
    \includegraphics[width=\linewidth]{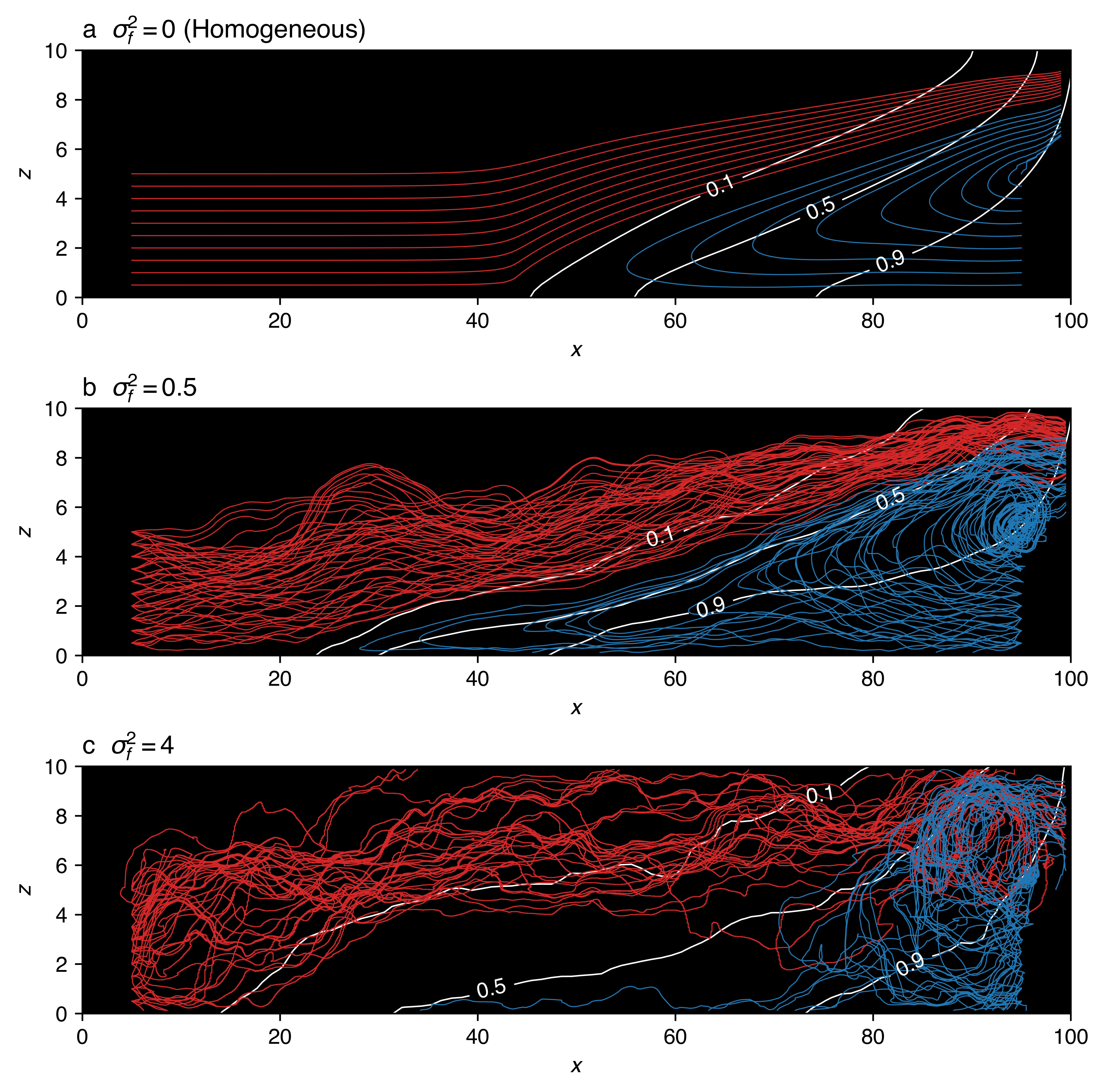}
    \includegraphics[width=\linewidth, pagebox=artbox]{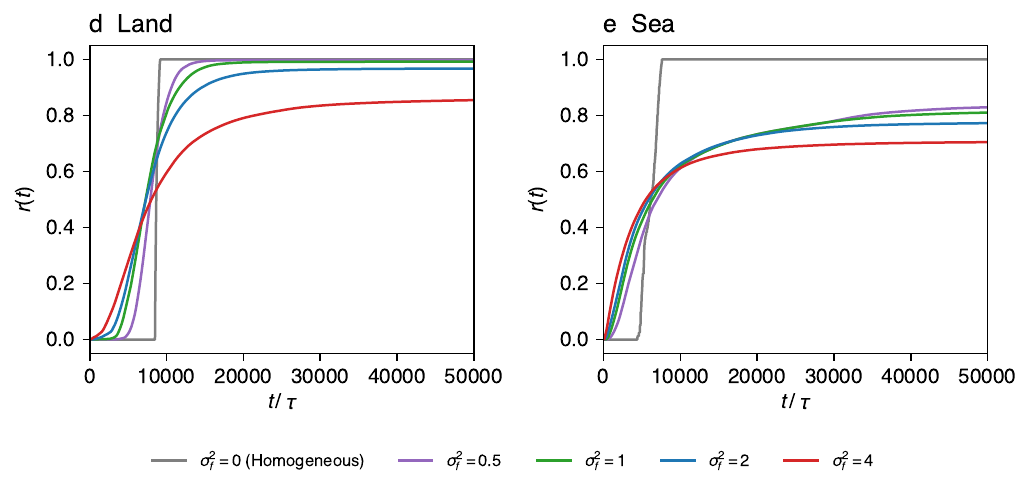}
    \caption{(a--c) Pathlines orthogonally projected onto $x-z$ plane, aligned with
      mean flow. Single realization is shown as example for each
      $\sigma^2_f$. Red and blue colours correspond to land and sea families, respectively. White contours denote normalised salinity. (d, e) Cumulative breakthrough curves at the coastal boundary ($x=0$) for
      particles originated from the (d) land ($x=5$ m) and (e) sea boundaries ($x=95$ m). Ensemble of 20 realizations for each $\sigma^2_f$. }
    \label{fig:pathline_btc}
  \end{figure}

Figures \ref{fig:pathline_btc}a--c illustrate pathlines belonging to particles from the sea and land families for
homogeneous and heterogeneous media up to time $t = 5 \times 10^4 \tau$
projected onto the $x-z$ plane. For the homogeneous
medium, the pathlines do not intersect and are fully segregated with the
particles from the land boundary flowing over the seawater body and
the seaward particles recirculating. This is different for heterogeneous media. In this case,
we observe pathline crossing and twisting that can give rise to the emergence of
periodic and chaotic orbits. For weakly heterogeneous media, the two families of
pathlines remain mostly segregated, while they intersect within the families. In
the recirculation zone, helical pathlines emerge, which manifest as closed lines
on the $x-z$ projection. For increasing heterogeneity, pathlines from the two
families cross each other and intertwine, that is, pathlines from the land
boundary penetrate the recirculation zone, and cross pathlines from the sea
boundary. The projections of pathlines in the recirculation zone describe closed
orbits, which indicate helical flow. These orbits can lead to a significant
increase in residence times or the trapping of solute particles. 

Figures~\ref{fig:pathline_btc}d and e show the cumulative distribution of residence
times for particles originating from the land and sea boundaries.
$R_L(t)$ and $R_S(t)$ increase over time, approaching a value of 1 as all the
particles discharge across the sea boundary in the homogeneous case. With
increasing heterogeneity, they evolve toward asymptotic values that are smaller
than $1$. For the homogeneous medium, the residence time distribution increases
abruptly almost as a step function for $R_L(t)$, which indicates little
dispersion along the pathlines. All the particles are recovered at a finite
time. With increasing heterogeneity, the width
of the residence time increases due to both early arrival due to preferential
paths and very long arrival times due to trapping in recirculating paths. The
numerical data indicate permanent trapping of up to 20\% of the
particles originating from the land and up to 35\% of the particles
originating from the sea boundary, which can be attributed to confined
pathlines.

\begin{figure}[htbp]
    \includegraphics[width=\linewidth, pagebox=artbox]{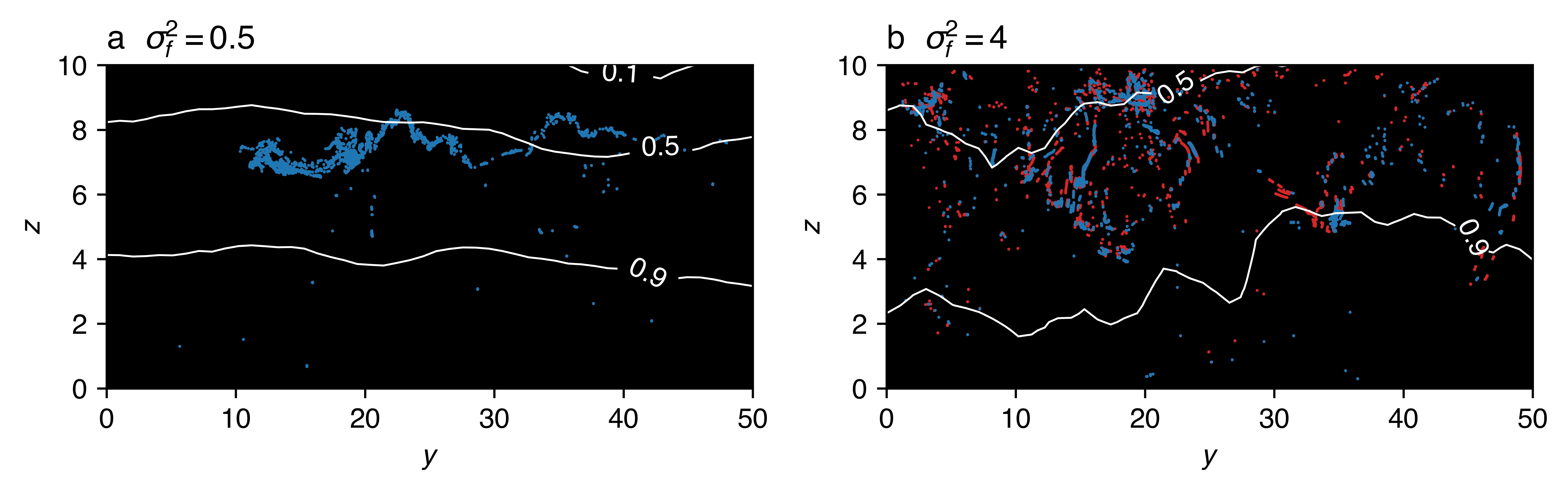}
    \includegraphics[width=\linewidth, pagebox=artbox]{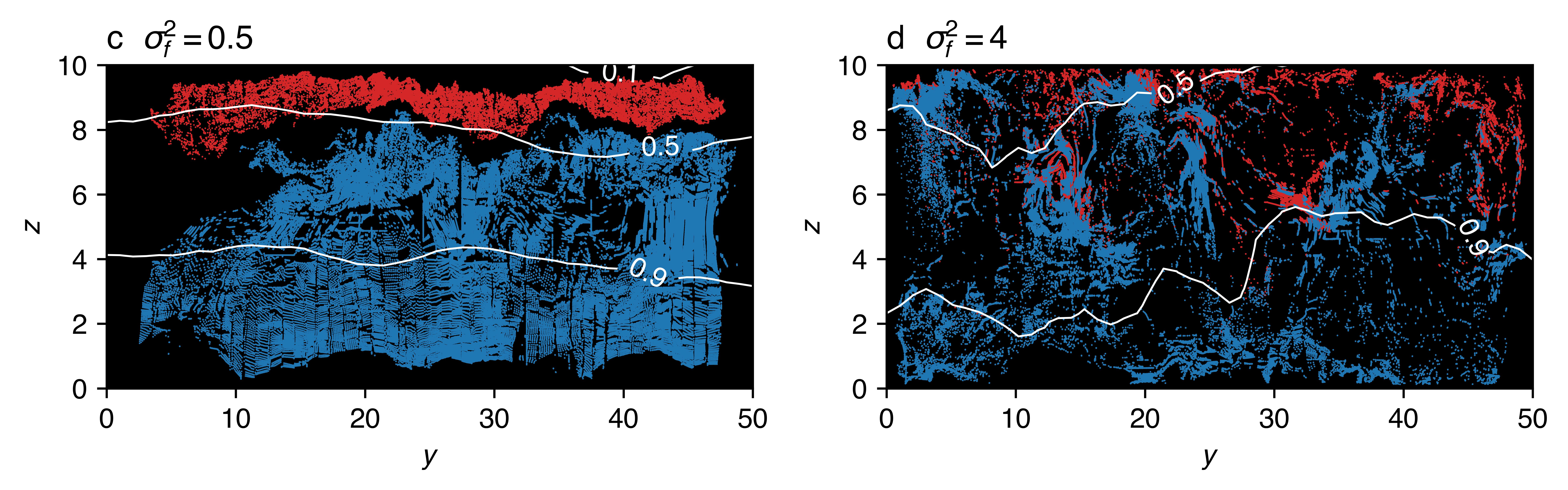}
    \caption{Poincar\'e sections at $x=90$ m, perpendicular to mean flow,
      for $t/\tau\leq5\times10^4$ for (a, b) particles with residence times above $3\times10^4$ s and (c, d) all particles. Single realization is shown as example for
      each $\sigma_f^2$. Colour scheme is identical to that in Figures
      \protect\ref{fig:pathline_btc}a--c.}
    \label{fig:poincare}
\end{figure}

Figures~\ref{fig:poincare}a and b show Poincar\'e sections based on the pathlines of
the particles whose residence times are above $3 \times 10^4$ s. The dots
indicate the intersection points of the particle paths at a plane at $x = 90$ m from
the sea boundary. We select a plane that intersects the recirculation zone close
to the sea boundary illustrated in Figures \ref{fig:pathline_btc} a--c.
For the weakly heterogeneous medium with $\sigma_f^2 = 0.5$,
the Poincar\'e section features a band of orbits of particles from the
sea boundary. With increasing heterogeneity, for $\sigma_f^2 = 4$, red and blue
dots are interspersed and dispersed across a
larger area than for $\sigma_f^2 = 0.5$. This indicates the potential for mixing
between water originating from the land and sea boundaries and contact between
these waters for long times. It also indicates the emergence of chaotic orbits
with increasing heterogeneity. 

Chaotic and periodic orbits effectively trap particles within the domain, mixing
particles and preventing their discharge into the sea. The retention of contaminants by the chaotic orbits in highly heterogeneous
aquifers imply long-term contamination risks of coastal aquifers. These
findings encompass not only contaminants but also nutrients from terrestrial
sources that are discharged to the sea as submarine groundwater discharge (SGD)
\cite{santos2021submarine, taniguchi2019submarine}. Such nutrient supply via SGD
affects marine biota by supporting primary productivity
\cite{waska2011submarine, waska2010differences, blanco2011estimation,
  adolf2019near}, whereas excess nutrient loadings can lead to eutrophication
\cite{hwang2005large, lee2009nutrient, kwon2017green, cho2019tracing}. In light of these studies, our findings imply the reduced transport of nutrients from
terrestrial sources to the sea via SGD, which might have multifaceted effects on
marine ecosystems, potentially limiting primary productivity or alleviating
eutrophication. Moreover, our results shed new light on the role of the
recirculation region as a hotspot for intense chemical and biological
reactions \cite{moore1999subterranean, heiss2017physical, robinson2009tidal,
  liu2018tidal}.

\subsection{Chaotic mixing between freshwater and seawater domains}
To shed further light on the mixing of particles originating from land and sea
boundaries, we analyze the distribution of intersection points on the Poincar\'e
sections at $x=90$ m. Figures \ref{fig:poincare}c and d show the intersection
points of all particles from the land and sea boundaries. For weak heterogeneity
($\sigma_f^2=0.5$), the dots from the land (red) and sea (blue) families
are well segregated, with the former confined to the freshwater domain and the
latter to the seawater domain. For increasing heterogeneity ($\sigma_f^2=4$), in
contrast, the two families intermix, with particles of the land family
penetrating the seawater domain and vice versa. These findings indicate that the
chaotic orbits, which emerge in the salinity transition zone near the coast,
enhance the mixing between contaminants originating in the freshwater and seawater
domains.
 
\begin{figure}[htbp]
    \includegraphics[width=\linewidth, pagebox=artbox]{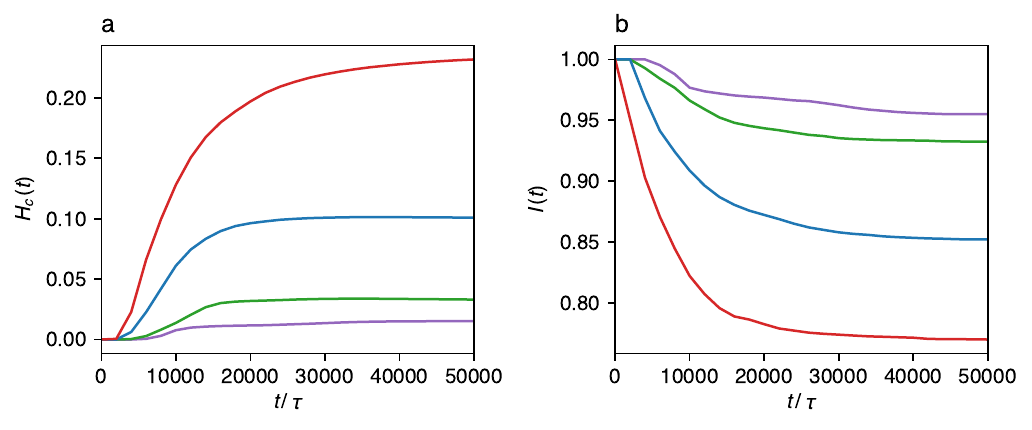}
    \includegraphics[width=\linewidth, pagebox=artbox]{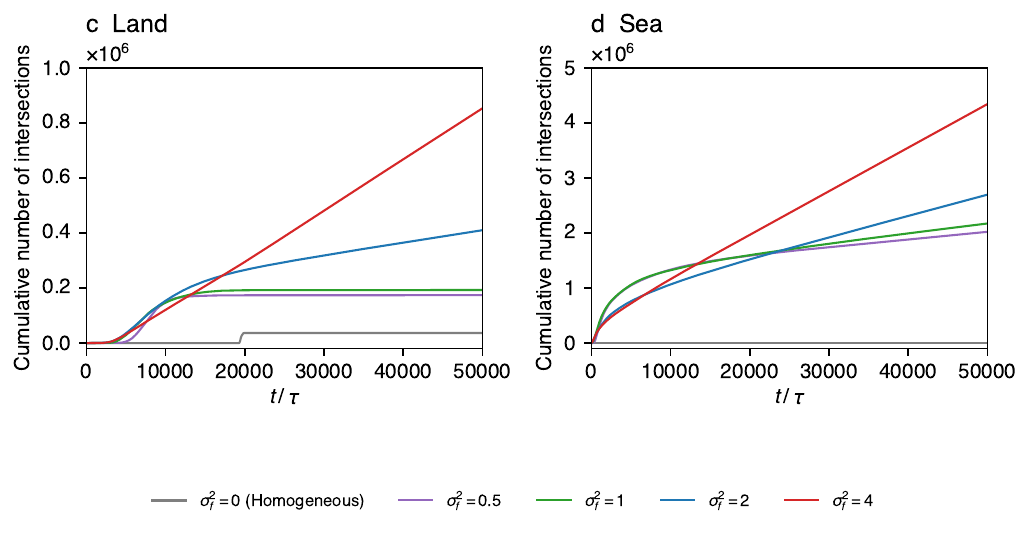}
    \caption{Temporal evolutions of (a) mixing entropy ($H_c(t)$), and (b) segregation intensity ($I(t)$), and (c, d) Cumulative number of intersection points for Poincar\'e sections in Figures~\ref{fig:poincare}c and d. Ensemble of 20 realisations for each $\sigma_f^2$.}
    \label{fig:mixing}
  \end{figure}

In order to quantify the quenching of dots belonging to the land and sea
families, we determine the temporal evolution of the mixing entropy $H_c(t)$ and
segregation intensity $I(t)$ between the two colours. Figures \ref{fig:mixing}a
and b show $H_c(t)$, which characterises the degree of mixing between the red
and blue dots, and $I(t)$, which quantifies the degree of segregation between
the two colours. $H_c(t)$ increases over time at a rate that increases with
$\sigma_f^2$ towards an asymptotic value that describes the asymptotic mixing
state between the two colours. Correspondingly, the segregation intensity $I(t)$
decreases over time at a rate that increases with increasing $\sigma_f^2$, see
Figure \ref{fig:mixing}b. These observations show that the mixing between the
land and sea families intensifies with increasing heterogeneity.

The existence and role of chaotic orbits for mixing is further corroborated by
the cumulative number of intersection points in the Poincar\'e map. If all particles
intersect only a finite number of times, the cumulative number converges to a
constant value. Likewise, if pathlines describe periodic orbits, the number of
intersections stabilizes at a constant value. This is what we observe in Figures~\ref{fig:mixing}c and d for
particles originating from the land boundary for homogeneous and weakly
heterogeneous media. For increasing heterogeneity, however, the number of
intersections increases linearly with time both for the particles from the sea
and land boundaries. This clearly indicates the existence of chaotic or
quasi-periodic orbits that intersect the plane after each period at a different point. Chaotic orbits are effective in enhancing mixing between contaminants originating from freshwater and seawater
domains because the intersection points disperse across the Poincar\'e
section. 

\section{Conclusion}
We investigate how the interplay of tidal forcing, buoyancy, and aquifer heterogeneity controls contaminant transport in coastal aquifers. 
Numerical simulations of variable density flow in three-dimensional heterogeneous aquifers provide evidence of the emergence of chaotic orbits and periodic orbits in the salinity transition zone near the coast with increasing heterogeneity. Closed orbits persistently trap contaminants within the aquifer, as seen in the distribution of solute residence times. 
The trapping of contaminants in closed orbits can pose long-term contamination risks in coastal aquifers with possible multifaceted impacts on marine ecosystems, potentially affecting nutrient-driven biological activity or eutrophication. On the other hand, chaotic orbits lead to enhanced mixing between salt and freshwater borne solute particles as evidenced by the behaviour of the mixing entropy and segregation intensity. These results elucidate the mechanisms that convert the saltwater-freshwater transition zone into a hotspot for chemical and biological activity \cite<e.g.>{moore1999subterranean, heiss2017physical, robinson2006driving, robinson2009tidal}.
Furthermore, our findings suggest that chaotic behaviours in flow and transport processes can also occur during infrequent events, such as storm surges \cite<e.g.>{tajima2023groundwater}, which are characterised by rapid changes in the flow field. These results underscore the critical importance of considering the interplay of tidal forcing, buoyancy and aquifer heterogeneity for the assessment of contaminant transport and chemical and biological activity in coastal aquifers.

\appendix
\section*{Appendix A. Variable-density flow}
We first note that fluid mass conservation is given by
\cite{mulligan2011tidal, guo2002user}
\begin{align}
\rho S_p \frac{\partial p(\vx,t)}{\partial t} + \phi \frac{d \rho(c)}{d c} \frac{\partial c(\vx,t)}{\partial t} + \nabla \cdot \mathbf \rho \mathbf  q(\vx,t) = 0,
\end{align}
where $S_p$ is specific storage in terms of pressure, $h(\vx,t)$ hydraulic head, $\mathbf
q(\vx,t)$ specific discharge, $\phi$ porosity and $c(\vx,t)$ is the volumetric
salt concentration. The fluid density $\rho(c)$ is assumed to depend linearly on
the salt concentration $c(\vx,t)$, that is, $\rho(c) = \rho_0 [1 + \epsilon
c(\vx,t)/c_s]$, where $c_s$ is the constant salt concentration in seawater, and
$\rho_0$ is the reference density of freshwater. Furthermore, we replace
pressure by hydraulic head $p(\vx,t) = h(\vx,t) \rho g$,
which gives  
\begin{align}
\rho S_f \frac{\partial h(\vx,t)}{\partial t} + \rho S_f  h(\vx,t) \frac{d \rho(c)}{d c}
\frac{\partial c(\vx,t)}{\partial t}  + \phi \frac{d \rho(c)}{d c}
\frac{\partial c(\vx,t)}{\partial t} + \nabla \cdot \mathbf \rho \mathbf  q(\vx,t) = 0,
\end{align}
where we defined the specific storativity $S_f = \rho g S_p$. In the paper, we
use the Oberbeck-Boussinesq approximation, which disregards
density variations in the continuity equation. As pointed out by
\citeA{guo2002user}, density effects in the continuity equation are
important only for strong density contrasts. Thus, in the following we set
\cite{Langevin2020}
\begin{align}
S_f \frac{\partial h(\vx,t)}{\partial t} + \nabla \cdot \mathbf q(\vx,t) = 0,
\end{align}
with $S_f = \rho_0 g S_p$.

\section*{Appendix B. Numerical scheme for particle tracking}
\label{app:numerical}
For the readers' convenience, this supporting information summarises the numerical scheme for particle tracking simulations used in this study. The scheme follows the methods for purely advective transport in a two-dimensional domain in previous studies \cite{tajima2024ted, tajima2024tide}, which is extended to three dimensions.

The Langevin equation is spatially distretised and numerically solved with the modified third-order Runge-Kutta method. The time step $\Delta t$ is divided into three semi-time steps, and the particle location at the $j$th semi-time step is approximated by \cite{Drummond1984, tajima2024ted}
\begin{equation}
    \vx_j = \vx_0 + \sum_{k=1}^{j}\alpha_{jk}\vu(\vx_{j-1})\Delta t \, \, (j = 1, 2, 3),
    \label{eq:RK}
\end{equation}
where $\vx_0 = \vx(t)$ and $\vx(t+\Delta t) = \vx_3$, and $\alpha_{jk}$ are the empirical parameters (see \citeA{Drummond1984} for the values).

The velocity within the cells is calculated using trilinear interpolation. Suppose the particle location $(x, y, z)$ is within a voxel cornered by eight lattice points $(x_0, y_0, z_0)$, $(x_0, y_0, z_1)$, $(x_0, y_1, z_1)$, $(x_0, y_1, z_0)$, $(x_1, y_0, z_0)$, $(x_1, y_0, z_1)$, $(x_1, y_1, z_1)$ and $(x_1, y_1, z_0)$, where $x_1 - x_0 = \Delta x$, $y_1 - y_0 = \Delta y$, and $z_1 - z_0 = \Delta y$. We write the flow velocity at each lattice point by $\vu_{000}, \vu_{001}, \vu_{011}, \vu_{010}, \vu_{100}, \vu_{101}, \vu_{111}$ and $\vu_{110}$, respectively. The flow velocity at the point $(x, y, z)$ is written as
\begin{align}
    \vu(x, y, z) = \vu_{000}(1-x_d)(1-y_d)(1-z_d) &+ \vu_{100}x_d(1-y_d)(1-z_d)\\
    + \vu_{010}(1-x_d)y_d(1-z_d) &+ \vu_{110}x_d y_d(1-z_d)\\
    + \vu_{001}(1-x_d)(1-y_d)z_d &+ \vu_{101}x_d(1-y_d)z_d\\
    + \vu_{011}(1-x_d)y_d z_d &+ \vu_{111}x_d y_d z_d,
\end{align}
where
\begin{align}
    x_d &= \frac{x - x_0}{\Delta x}\\
    y_d &= \frac{y - y_0}{\Delta y}\\
    z_d &= \frac{z - z_0}{\Delta z}.
\end{align}

To mitigate numerical errors, we implemented an adoptive time-step control using the step-doubling scheme \cite{Press1992}. The initial time step interval is set to \cite{deDreuzy2007, tajima2024ted}
\begin{equation}
    \Delta t^{0} = 0.1 \max\left[\frac{\Delta x}{\max(u_x)}, \, \frac{\Delta y}{\max(u_y)} \, \frac{\Delta z}{\max(u_z)}\right].
\end{equation}
After calculating the particle location at a time step $t^n$ with Equation (\ref{eq:RK}), the next time-step interval $\Delta t^{n+1} = t^{n+2} - t^{n+1}$ is determined by the following procedure \cite{Diersch2013, tajima2024ted}: 
\begin{enumerate}
    \item Calculate the particle location $\mathbf{x}^{n+1}_m$ for a full time step $\Delta t$ with Equation (\ref{eq:RK}).
    \item Calculate the location of the particles $\mathbf{x}^{n+1/2}_m$ for a half-step $\Delta t / 2$ with Equation (\ref{eq:RK}). Then, $\widetilde{\mathbf{x}^{n+1}_m}$ (the location of the particles after the two half steps) is calculated similarly for the remaining half step $\Delta t / 2$ with $\mathbf{x}^{n+1/2}_m$.
    \item Calculate the maximum relative difference $d^{n+1}$ between $\mathbf{x}^{n+1}_m$ and $\widetilde{\mathbf{x}^{n+1}_m}$ by
    \begin{equation}
        d^{n+1} = \max \left(\frac{|x^{n+1}_m - \widetilde{x^{n+1}_m}|}{|x^n| + |u_x^n \Delta t|}, \, \frac{|y^{n+1}_m - \widetilde{y^{n+1}_m}|}{|y^n| + |u_y^n \Delta t|}, \, \frac{|z^{n+1}_m - \widetilde{z^{n+1}_m}|}{|z^n| + |u_z^n \Delta t|} \right).
    \end{equation}
    \item Determine $\Delta t^{n+1}$ by
    \begin{equation}
    \Delta t^{n+1} =
    \left\{
    \begin{array}{ll}
            0.9 \Delta t ^n \left(\frac{\varepsilon}{d^{n+1}}\right)^{\frac{1}{4}} & \textrm{if} \,\, d^{n+1} \leq \varepsilon \\
            0.9 \Delta t ^n \left(\frac{\varepsilon}{d^{n+1}}\right)^{\frac{1}{3}} & \textrm{if} \,\, d^{n+1} > \varepsilon
    \end{array}
    \right.
    ,
    \end{equation}
    where $\varepsilon$ ($=10^{-3}$ in this study) is the pre-defined maximum error.       
\end{enumerate}

\section*{Appendix C. Parameters for numerical simulations}
Table \ref{tab:parameters} summarises the parameters used in the numerical simulations, including
the density-dependent flow and transport simulations with MODFLOW 6
\cite{Langevin2017} and the particle tracking simulations for purely advective
contaminant transport.
\begin{table}
    \caption{Parameters for numerical simulations}
    \centering
    \begin{tabular}{l c c}
    \hline
    Parameter  & Value & Unit  \\
    \hline
    Domain size ($L_x \times L_y \times L_z$) & $100\times50\times10$ & m\\
    Geometric mean hydraulic conductivity ($K_G$) & 3.5 & m d$^{-1}$\\
    Log-conductivity variance ($\sigma_f^2$) & 0.5, 1, 2, 4 & --\\
    Correlation lengths ($\lambda_x \times \lambda_y \times \lambda_z$) & $5\times5\times2$ & m\\
    Grid sizes ($\Delta_x \times \Delta_y \times \Delta_z$) & $1\times1\times0.25$ & m\\
    Amplitude ($A$) & 0.5 & m\\
    Period ($\tau$) & 0.5 & d\\
    Specific storage ($S_s$) & $1\times10^{-2}$ & m$^{-1}$\\
    Porosity & 0.25 & --\\
    Terrestrial flux ($q_\mathrm{f}$) & $1\times10^{-3}$ & m d$^{-1}$\\
    Longitudinal dispersivity & 0.5 & m\\
    Transverse dispersivity & 0.05 & m\\
    Molecular diffusivity\tablefootnote{Only applied to the density-dependent flow and transport simulations with MODFLOW. In the particle tracking simulations for a contaminant, molecular diffusivity is zero (purely advective).} & $8.64\times10^{-5}$ & m$^2$ d$^{-1}$\\
    Freshwater density & $1.0000\times10^{-3}$ & kg m$^{-3}$\\
    Seawater density & $1.0245\times10^{-3}$ & kg m$^{-3}$\\
    Gravitational acceleration & 9.8 & m s$^{-2}$\\
    Viscosity & $1.124\times10^{-3}$ & Pa s\\
    \hline
    \end{tabular}
    \label{tab:parameters}
\end{table}

For calculating and illustrating travel distances, breakthrough curves, and Poincar\'e sections,
a total of $36,531$ particles are seeded at $0.1$ and $0.05$ m intervals in $y-$ and
$z-$directions. For calculating the mixing entropy and segregation intensity at
a total of $181,101$ particles are seeded at 0.05 and 0.02 m intervals in each direction.
\acknowledgments
ST acknowledges insightful suggestions from Niels Hartog. MD acknowledges funding by the European Union (ERC, KARST, 101071836). Views and opinions expressed are, however, those of the authors only, and do not necessarily reflect those of the European Union or the European Research Council Executive Agency. Neither the European Union nor the granting authority can be held responsible for them. The authors declare no competing interests.

\bibliography{tide-3d}

\end{document}